\documentclass[12pt,letter]{article}
\usepackage[margin = 1in]{geometry}
\usepackage{indentfirst}
\usepackage{latexsym,graphicx}
\begin{document}

\begin{center}
\LARGE\textbf{Negative frequency communication}
\end{center}

\begin{center}
Fanping DU\\
Email: dufanping@hotmail.com
\end{center}

\begin{center}
Qing Huo Liu\\
Department of Electrical and Computer Engineering\\
Duke University\\
Email: Qing.Liu@duke.edu
\end{center}

\begin{abstract}
Spectrum is the most valuable resource in communication system, but unfortunately, so far, a half of the spectrum has been wasted. In this paper, we will see that the negative frequency not only has a physical meaning but also can be used in communication. In fact, the complete description of a frequency signal is a rotating complex-frequency signal, in a complete description, positive and negative frequency signals are two distinguishable and independent frequency signals, they can carry different information. But the current carrier modulation and demodulation do not distinguish positive and negative frequencies, so half of the spectrum resources and signal energy are wasted. The complex-carrier modulation and demodulation, proposed by this paper, use the complex-frequency signal as a carrier signal, the negative and positive frequency can carry different information, so the spectrum resources are fully used, the signal energy carried by complex-carrier modulation is focused on a certain band, so the signal energy will not be lost by the complex-carrier demodulation.
\end{abstract}

\normalsize
\section{Introduction}
According to Shannon formula:
\begin{equation}
\label{eq1}
C=W\ast \log_2^{(1+S/N)}
\end{equation}
where $C$ is the channel capacity, $W$ is the channel bandwidth, $S$ is the signal power, and $N$ is the noise power. We can see that the most effective way to increase the channel capacity is increasing bandwidth, and enhance the signal-to-noise ratio can also increase the channel capacity.

\begin{figure}
\includegraphics[width=5in]{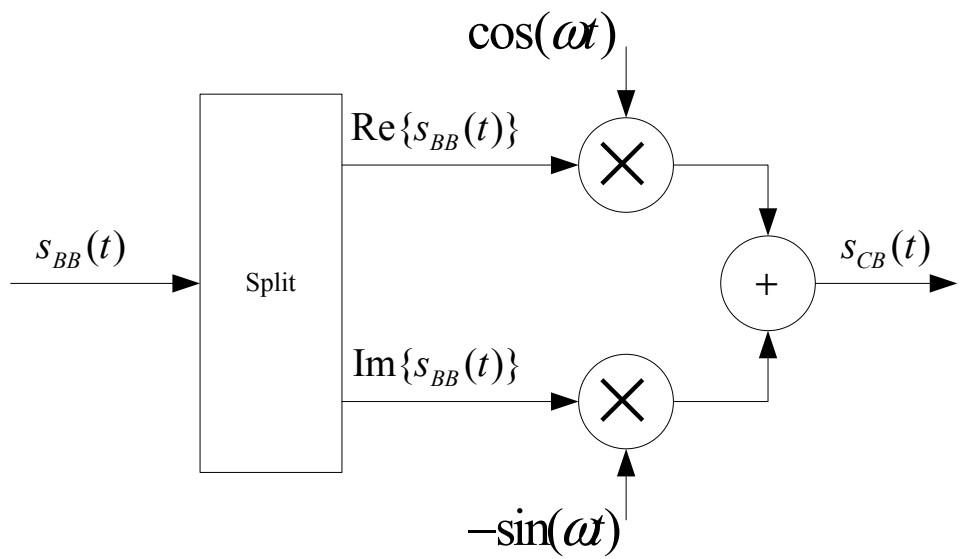}\\
\caption{Principle of real-carrier modulation.}\label{fig1}
\end{figure}

In current communication systems, to take full advantage of the spectrum, carrier modulation technology is used. The principle of current carrier modulation is shown in Fig. \ref{fig1}, the real part of the baseband complex signal multiplied by $cos(\omega t)$ and the imaginary part multiplied by $-sin(\omega t)$, then add, as the following formula:
\begin{equation}
\label{eq2}
s_{CB} (t)=Re\{s_{BB} (t)\ast e^{i\omega t}\}
\end{equation}
where $s_{CB} (t)$ is the modulation signal, $e^{i\omega t}$ is the carrier signal, $s_{BB} (t)$ is the baseband complex signal, $Re$ is to take the real part. Obviously in the current carrier modulation, although the baseband signal is complex, the carrier signal is complex, but the modulation signal is taken the real part, so a real signal is sent. In this paper it is called real-carrier modulation.

In fact, the real-carrier modulation technology wastes a half of the spectrum resources and the signal energy, that is because of the incorrect understanding and using of negative frequency. So far, in all current communication systems, including wireless, cable and fiber the available bandwidth is defined in the positive spectrum. For example, the bandwidth defined in the latest version of LTE protocol, see \cite{bib2}, negative spectrum is not used. So to make full use of spectrum resources, we must first rectify the understanding of the negative frequency.

\section{Complex-frequency signal}
\begin{figure}
\includegraphics[width=5in]{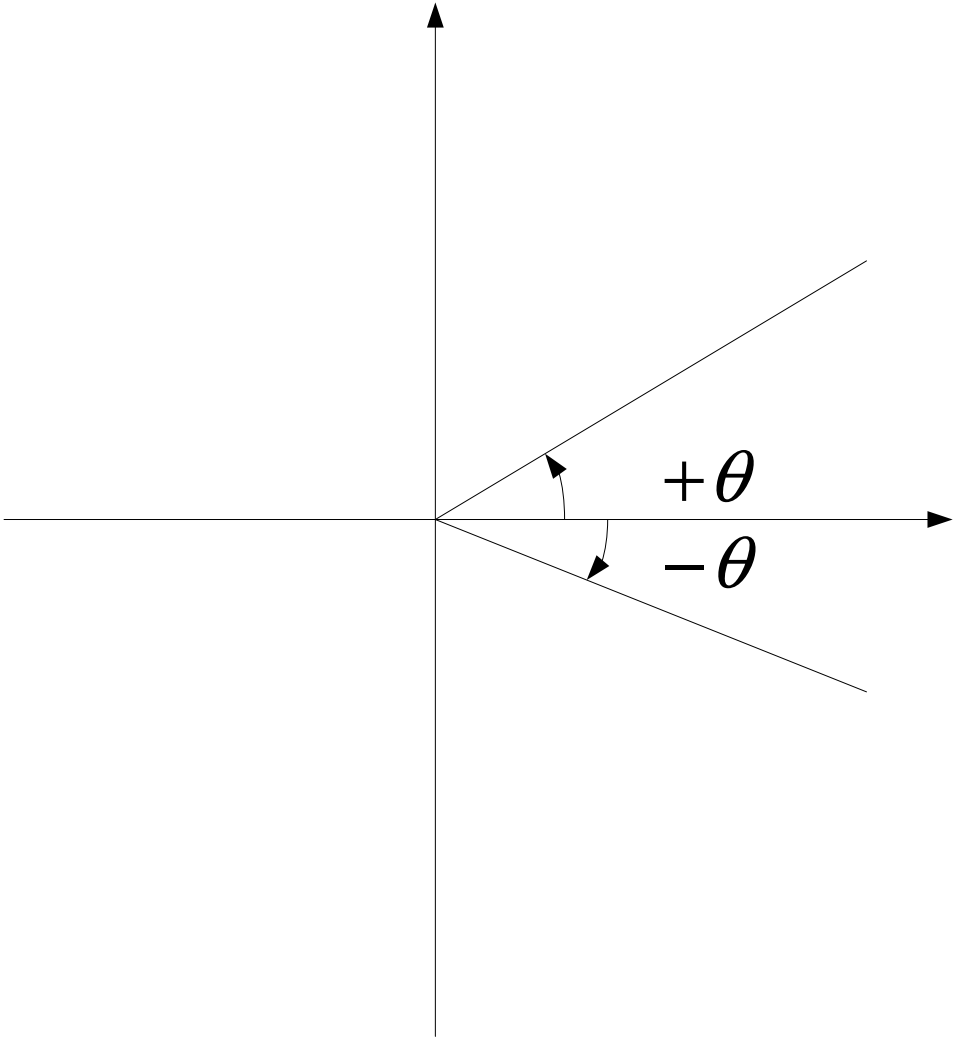}\\
\caption{Definition of the angle.}\label{fig2}
\end{figure}

The negative frequency has a physical meaning, see \cite{bib1}. As shown in Fig. \ref{fig2}, we define the angle of counterclockwise rotation as $+\theta$, the angle of clockwise rotation as $-\theta$, so the definition of angular frequency is:
\begin{equation}
\label{eq3}
\omega  = \frac{{d\theta(t)}}{{dt}}
\end{equation}
\begin{equation}
\label{eq4}
- \omega  = \frac{{d(-\theta(t))}}{{dt}}
\end{equation}
By (\ref{eq3}), (\ref{eq4}) we can see that the positive angular frequency corresponds to the speed of counterclockwise rotation, the negative angular frequency corresponds to the speed of clockwise rotation, so ``negative angular frequency'' does not due to the ``negative time'', but due to ``negative angle'', positive and negative frequency represent two different directions of rotation in a plane, there are two kinds of frequencies is essentially because of the angle is defined in a plane, a plane has two sides.

With the understanding of the physical meaning of negative frequency, then how to distinguish between positive and negative frequencies, or how to describe these two directions of rotations in a plane? That is the Euler formula:
\begin{equation}
\label{eq5}
{e^{ \pm i\omega t}} = \cos (\omega t) \pm i*\sin (\omega t)
\end{equation}

\begin{figure}
\includegraphics[width=5in]{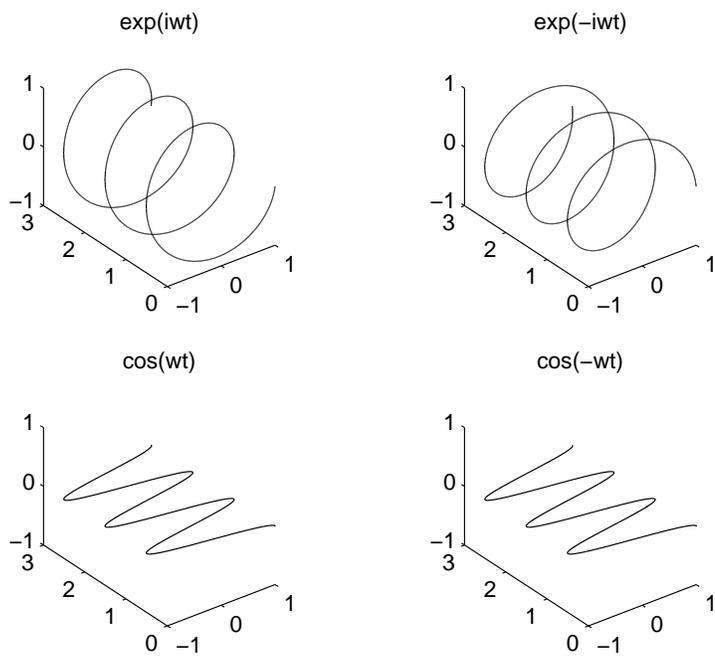}\\
\caption{Positive and negative frequency signals and their projections.}\label{fig3}
\end{figure}

As shown in Fig. \ref{fig3}, ${e^{ - i\omega t}}$ and ${e^{i\omega t}}$ are two kinds of complex-frequency signals, corresponding to the negative and positive frequency signals. Although in the ``time-complex'' direct product space, it is easy to distinguish positive and negative frequency signals, but in ``time-real'' direct product space, the projection of positive and negative frequency signals are the same real-frequency signal $cos(\omega t)$, that is:
\begin{equation}
\label{eq6}
Re\{ {e^{ - i\omega t}}\}  = Re\{ {e^{i\omega t}}\}  = \cos (\omega t)
\end{equation}
So when we see a real-frequency signal, we can not distinguish it between positive or negative frequency signals from its projection, the probability of the frequency signal being negative or positive is equal, it is 1/2, that is:
\begin{equation}
\label{eq7}
\cos (\omega t) = ({e^{ - i\omega t}} + {e^{i\omega t}})/2
\end{equation}
Similarly:
\begin{equation}
\label{eq8}
\sin (\omega t) = i*({e^{ - i\omega t}} - {e^{i\omega t}})/2
\end{equation}
Therefore, a real-frequency signal $cos(\omega t)$ or $sin(\omega t)$ is not complete, the complete description of a frequency signal must be a rotating complex-frequency signal ${e^{ \pm i\omega t}}$, in complete description, the negative frequency signal ${e^{ \- i\omega t}}$ and positive frequency signal ${e^{ \i\omega t}}$ are two distinguishable and independent frequency signals, they can carry different information.

In this paper, we define the positive frequency, which direction of rotation meet the right hand rule as the R-frequency. We define the negative frequency, which direction of rotation meet the left hand rule as the L-frequency. Unless otherwise cited, this paper will use the terms L-frequency and R-frequency, L-band and R-band instead of the positive frequency and negative frequency, positive band and negative band.

\section{Real-carrier modulation and demodulation}
In this section, we will see that the current real-carrier modulation technology occupies all the L-band and R-band in the spectrum, so a half of the spectrum resources are wasted, and the current real-carrier demodulation technology only receives one of the L-band or R-band, so a half of the signal energy is lost.

\subsection{Real-carrier modulation}
The principle of current real-carrier modulation is as formula (\ref{eq2}). It is modulated by the R-frequency signal ${e^{ \i\omega t}}$, according to the formula (\ref{eq7}) and (\ref{eq8}):
\begin{small}
\begin{equation}
\label{eq9}
\begin{array}{l}
{s_{CB}}(t) \\
=Re\{ {s_{BB}}(t)*{e^{i\omega t}}\} \\
=Re\{ {s_{BB}}(t)\} *\cos (\omega t) - Im\{ {s_{BB}}(t)\} *\sin (\omega t) \\
=Re\{ {s_{BB}}(t)\} *({e^{ - i\omega t}} + {e^{i\omega t}})/2 - Im\{ {s_{BB}}(t)\} *i*({e^{ - i\omega t}} - {e^{i\omega t}})/2 \\
=(Re\{ {s_{BB}}(t)\}  - Im\{ {s_{BB}}(t)\} *i)*{e^{ - i\omega t}}/2 + (Re\{ {s_{BB}}(t)\}  + Im\{ {s_{BB}}(t)\} *i)*{e^{i\omega t}}/2 \\
={s_{BB}}{(t)^*}*{e^{ - i\omega t}}/2 + {s_{BB}}(t)*{e^{i\omega t}}/2 \\
\end{array}
\end{equation}
\end{small}
where ${s_{CB}}(t)$ is the modulation signal, ${s_{BB}}(t)$ is the baseband signal, ${s_{BB}}(t)^*$ is the conjugated baseband signal, ${e^{i\omega t}}$ is the R-frequency signal. According to the multiplication of two signals in time domain is equivalent to their convolution in frequency domain, it can be seen from the above equation that the real-carrier modulation will move the baseband to the R-band and L-band, and the signal energy is divided equally.

Therefore, after being modulated by the R-frequency signal ${e^{i\omega t}}$ and taking the real part, the R-band signal is the same as the baseband signal, and the amplitude is a half, the L-band signal is conjugated symmetry to the baseband signal, and the amplitude is also a half.

Similarly, if it is modulated by the L-frequency signal ${e^{- i\omega t}}$:
\begin{equation}
\label{eq10}
{s_{CB}}(t) = Re\{ {s_{BB}}(t)*{e^{ - i\omega t}}\} = {s_{BB}}(t)*{e^{ - i\omega t}}/2 + {s_{BB}}{(t)^*}*{e^{i\omega t}}/2
\end{equation}
where ${s_{CB}}(t)$ is the modulation signal, ${s_{BB}}(t)$ is the baseband signal, ${s_{BB}}(t)^*$ is the conjugated baseband signal, ${e^{- i\omega t}}$ is the L-frequency signal.

Therefore, after modulated by the L-frequency signal ${e^{- i\omega t}}$ and take the real part, the L-band signal is the same as the baseband signal, and the amplitude is a half, the R-band signal is conjugated symmetry to the baseband signal, and the amplitude is also a half.

In summary, the real-carrier modulation occupies all the L-band and R-band, and the information on the L-band and R-band are conjugated symmetric, not independent, and the signal energy is a half on each side.

Furthermore, currently people regard one of the bands as a ``image frequency component'', as if it is not reality, in fact, the ``image frequency'' is caused by using the incomplete real-frequency signal as the carrier signal.

\begin{figure}
\includegraphics[width=5in]{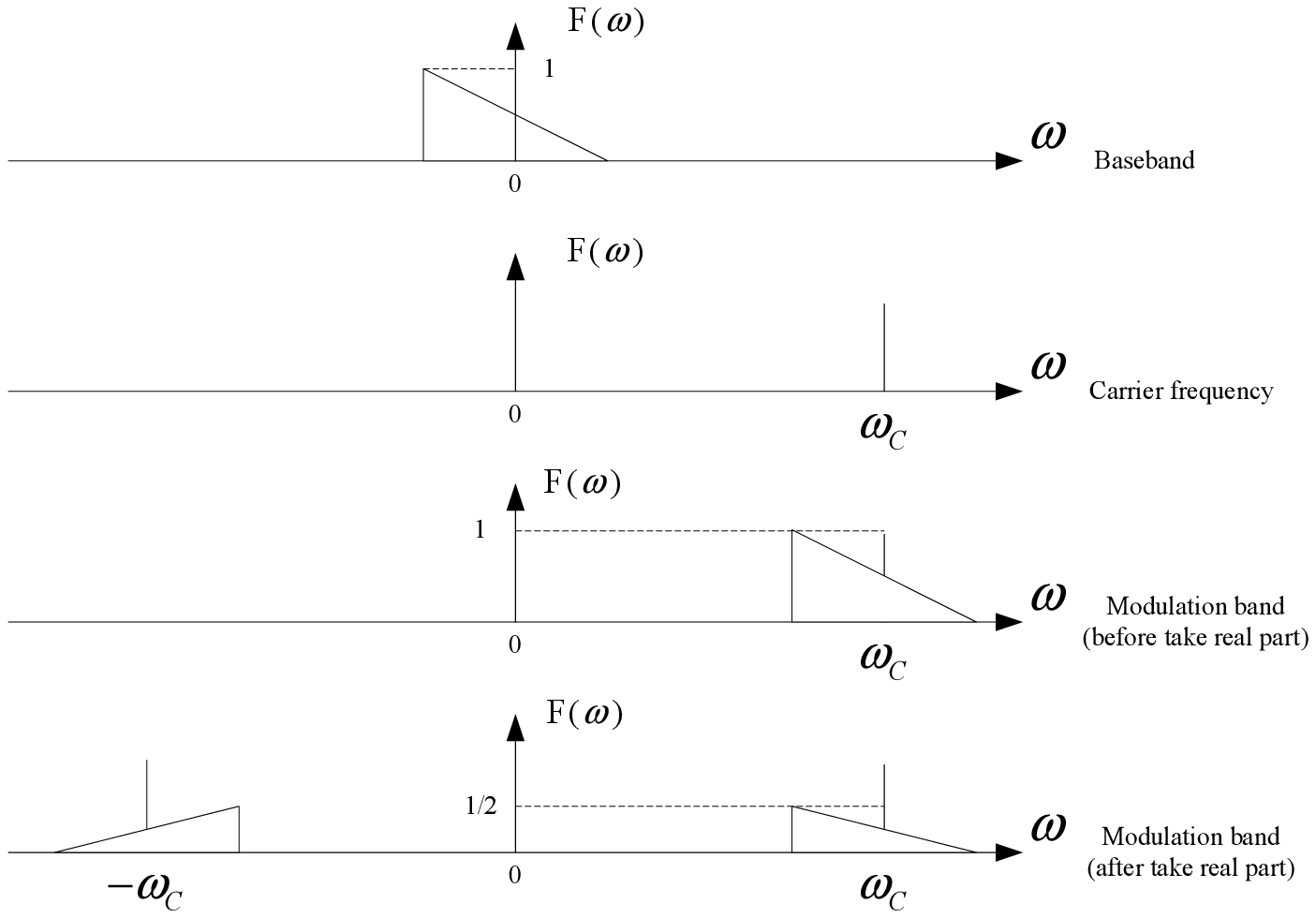}\\
\caption{Band move of the real-carrier modulation.}\label{fig4}
\end{figure}

A demo of band move of the real-carrier modulation is shown in Fig. \ref{fig4}, modulated by the R-frequency signal, amplitude spectrum.

\subsection{Real-carrier demodulation}
The current real-carrier demodulation also assumed to receive a real signal, so it is dealt as a real signal. As the real-carrier modulation does not distinguish L-frequency and R-frequency, so the modulation signal maybe modulated by the L-frequency signal or the R-frequency signal, here assume the modulation signal is modulated by the R-frequency signal, as formula (\ref{eq9}).

If demodulated by the L-frequency signal ${e^{ - i\omega t}}$, according to the formula (\ref{eq9}):
\begin{equation}
\label{eq11}
\begin{array}{l}
{s_{RBB}}(t) = {s_{CB}}(t)*{e^{ - i\omega t}} \\
= ({s_{BB}}{(t)^*}*{e^{ - i\omega t}}/2 + {s_{BB}}(t)*{e^{i\omega t}}/2)*{e^{ - i\omega t}} \\
= {s_{BB}}{(t)^*}*{e^{ - 2*i\omega t}}/2 + {s_{BB}}(t)/2 \\
\end{array}
\end{equation}
where ${s_{RBB}}(t)$ is the demodulation signal, ${s_{CB}}(t)$ is the modulation signal, ${s_{BB}}(t)$ is the baseband signal, ${s_{BB}}(t)^*$ is the conjugated baseband signal, ${e^{ - i\omega t}}$ is the L-frequency signal. After demodulation, the L-band ${s_{BB}}{(t)^*}/2$ is moved to two times far away and the R-band ${s_{BB}}(t)/2$ is moved to the baseband, after a low-pass filter, the L-band signal energy is discarded.

Similarly, if demodulated by the R-frequency signal ${e^{i\omega t}}$, according to the formula (\ref{eq9}):
\begin{equation}
\label{eq12}
\begin{array}{l}
{s_{RBB}}(t) = {s_{CB}}(t)*{e^{i\omega t}} \\
= ({s_{BB}}{(t)^*}*{e^{ - i\omega t}}/2 + {s_{BB}}(t)*{e^{i\omega t}}/2)*{e^{i\omega t}} \\
= {s_{BB}}(t)*{e^{2*i\omega t}}/2 + {s_{BB}}{(t)^*}/2 \\
\end{array}
\end{equation}
where ${s_{RBB}}(t)$ is the demodulation signal, ${s_{CB}}(t)$ is the modulation signal, ${s_{BB}}(t)$ is the baseband signal, ${s_{BB}}(t)^*$ is the conjugated baseband signal, ${e^{i\omega t}}$ is the R-frequency signal. After demodulation, the R-band ${s_{BB}}(t)/2$ is moved to two times far away and the L-band ${s_{BB}}{(t)^*}/2$ is moved to the baseband, after a low-pass filter, the R-band signal energy is discarded. But the reserved L-band signal is conjugated to original signal.

Although information on the L-band and R-band is conjugated, but filtering out one of them will lost a half of signal energy.

\begin{figure}
\includegraphics[width=5in]{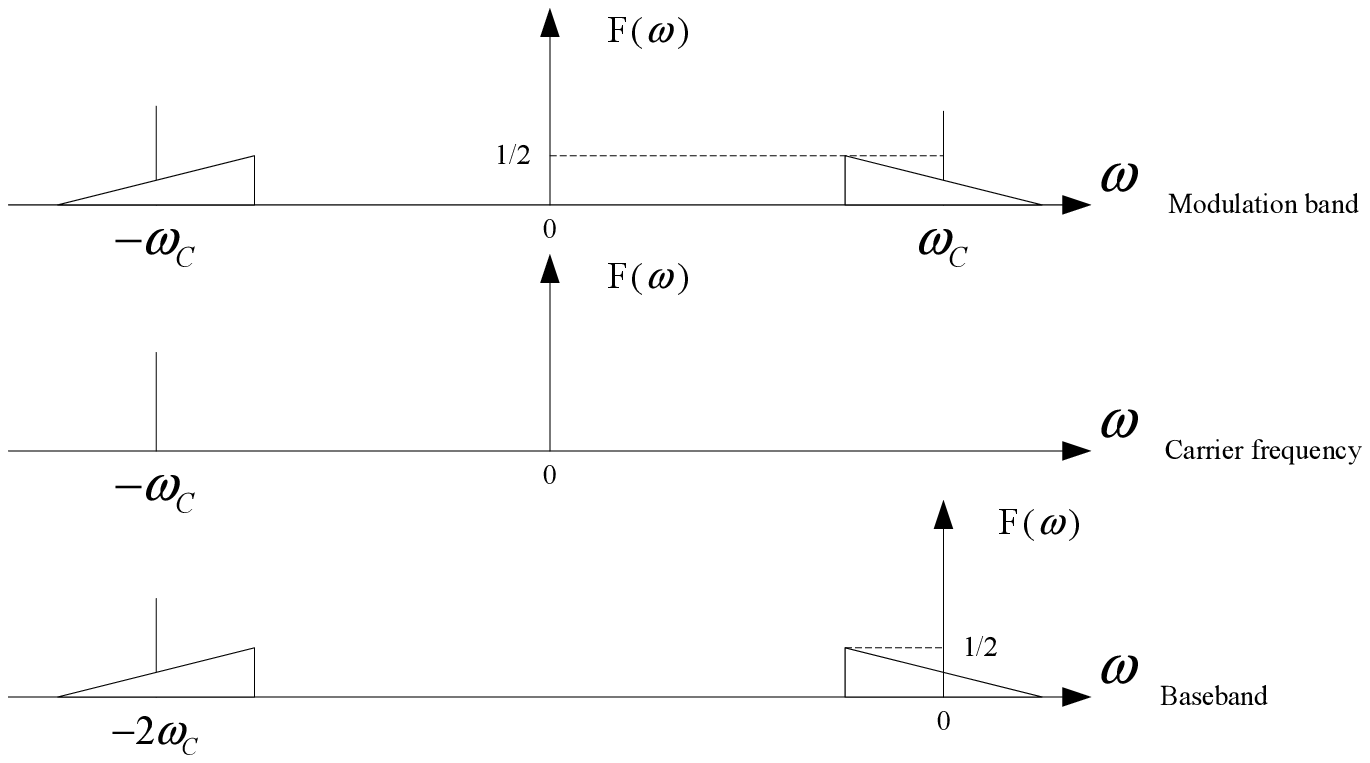}\\
\caption{Band move of the real-carrier demodulation.}\label{fig5}
\end{figure}

A demo of band move of the real-carrier demodulation is shown in Fig. \ref{fig5}, demodulated by the L-frequency signal, amplitude spectrum.

\section{Complex-carrier modulation and demodulation}
As mentioned earlier, the complete description of a frequency signal is a rotating complex-frequency signal, in a complete descriptions, the L-frequency signal ${e^{-i\omega t}}$ and R-frequency signal ${e^{i\omega t}}$ are two distinguishable and independent frequency signals, so they can carry different information. Therefore, we can modulate the baseband signals by the L-frequency or R-frequency signal, in order to distinguish from the real-carrier modulation, it is called complex-carrier modulation in this paper. Because there are two kinds of frequencies, so there are two kinds of complex-carrier modulations, in this paper the modulation using the L-frequency signal is called L-complex modulation, the modulation using the R-frequency signal is called R-complex modulation.

In this section, we will see that compared with the real-carrier modulation, the complex-carrier modulation uses the distinguishable and independent L-frequency signal ${e^{-i\omega t}}$ and R-frequency signal ${e^{i\omega t}}$ as the carrier signal, they can carry different information, so the spectrum resources are full used, the signal energy carried by complex-carrier modulation is focused on a certain band, so the signal energy will not be lost by the complex-carrier demodulation.

\subsection{Complex-carrier modulation}
There are two kinds of complex-carrier modulation, L-complex modulation and R-complex modulation.

The principle of L-complex modulation is as the following formula:
\begin{scriptsize}
\begin{equation}
\label{eq13}
\begin{array}{l}
{s_{CB}}(t) = {s_{BB}}(t)*{e^{ - i\omega t}} \\
= (Re\{ {s_{BB}}(t)\}  + i*Im\{ {s_{BB}}(t)\} )*(\cos (\omega t) - i*\sin (\omega t)) \\
= (Re\{ {s_{BB}}(t)\} *\cos (\omega t) + Im\{ {s_{BB}}(t)\} *\sin (\omega t)) + i*(Im\{ {s_{BB}}(t)\} *\cos (\omega t) - Re\{ {s_{BB}}(t)\} *\sin (\omega t)) \\
\end{array}
\end{equation}
\end{scriptsize}
where ${s_{CB}}(t)$ is the complex-carrier modulation signal, ${s_{BB}}(t)$ is baseband signal, ${e^{ - i\omega t}}$ is the L-frequency signal.

Similarly, the principle of R-complex modulation is as the following formula:
\begin{scriptsize}
\begin{equation}
\label{eq14}
\begin{array}{l}
{s_{CB}}(t) = {s_{BB}}(t)*{e^{i\omega t}} \\
= (Re\{ {s_{BB}}(t)\}  + i*Im\{ {s_{BB}}(t)\} )*(\cos (\omega t) + i*\sin (\omega t)) \\
= (Re\{ {s_{BB}}(t)\} *\cos (\omega t) - Im\{ {s_{BB}}(t)\} *\sin (\omega t)) + i*(Im\{ {s_{BB}}(t)\} *\cos (\omega t) + Re\{ {s_{BB}}(t)\} *\sin (\omega t)) \\
\end{array}
\end{equation}
\end{scriptsize}
where ${s_{CB}}(t)$ is the complex-carrier modulation signal, ${s_{BB}}(t)$ is baseband signal, ${e^{i\omega t}}$ is the R-frequency signal.

\begin{figure}
\includegraphics[width=5in]{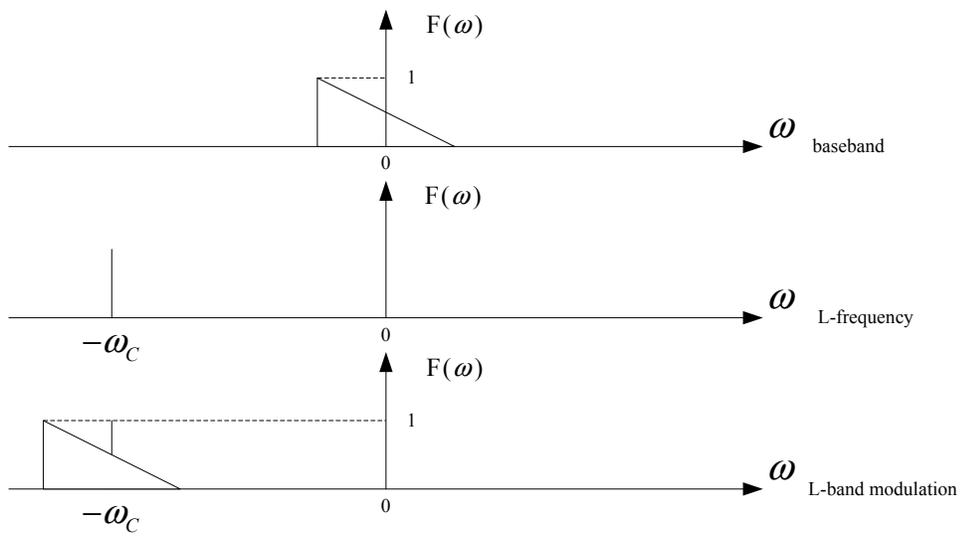}\\
\caption{Band move of the L-complex modulation.}\label{fig6}
\end{figure}

Band move of the L-complex modulation is shown in Fig. \ref{fig6}. Band move of the R-complex modulation is similar and omitted.

\begin{figure}
\includegraphics[width=5in]{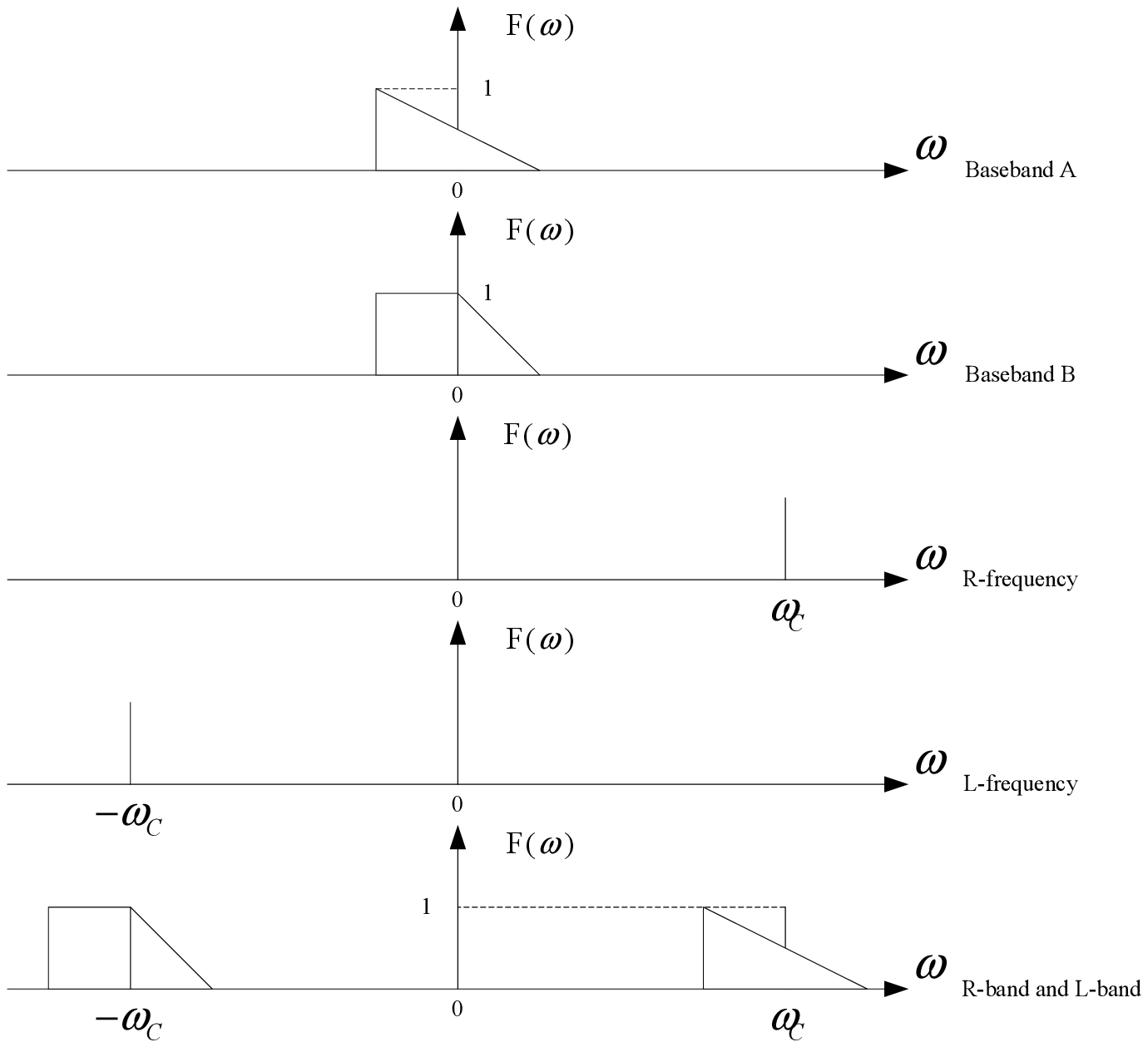}\\
\caption{Band move of modulating two different information.}\label{fig7}
\end{figure}

The L-frequency and R-frequency are two distinguishable and independent frequency signals, so they can carry different information. The band move of carrying two different information is shown in Fig. \ref{fig7}, the baseband A and B are moved to L-band and R-band separately.

\begin{figure}
\includegraphics[width=5in]{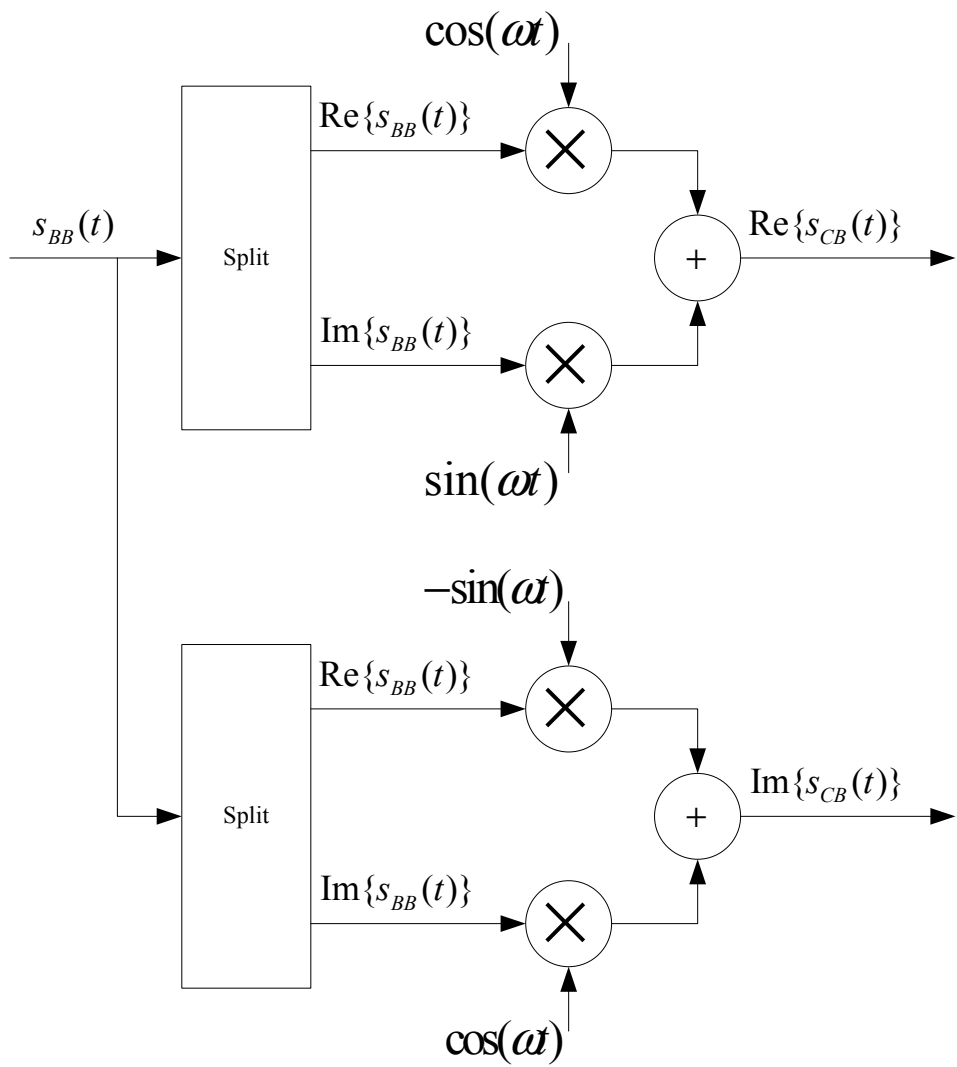}\\
\caption{Principle of L-complex modulation.}\label{fig8}
\end{figure}

According to the formula (\ref{eq13}) the principle of L-complex modulation is shown in Fig. \ref{fig8}:

It can be seen from the figure, the real part and the imaginary part of the L-complex modulation signal are modulated separately, they are two orthogonal signals in space, so the L-complex modulation signals in the transmission medium are the left rotating complex signals.

The principle of R-complex modulation is similar and omitted.

\subsection{Complex-carrier demodulation}
In essence, the principle of complex-carrier modulation and demodulation are the same, they are simply band moves, but in opposite directions. So the L-complex modulation signal is demodulated by the R-frequency signal, and the R-complex modulation signal is demodulated by the L-frequency signal.

\begin{figure}
\includegraphics[width=5in]{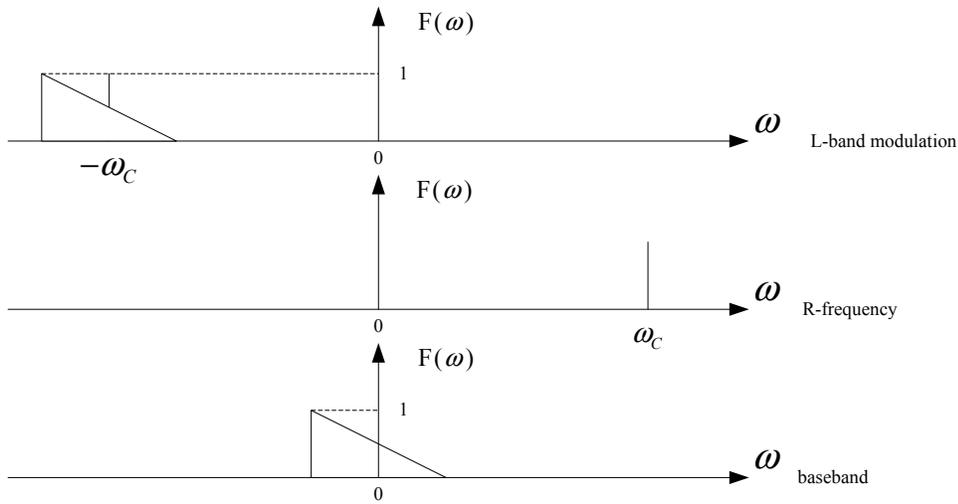}\\
\caption{Band move of the demodulation for the L-complex modulation signal.}\label{fig9}
\end{figure}

The L-complex modulation signal is demodulated by the R-frequency signal as the following formula:
\begin{equation}
\label{eq15}
{s_{RBB}}(t) = {s_{CB}}(t)*{e^{i\omega t}} = ({s_{BB}}(t)*{e^{ - i\omega t}})*{e^{i\omega t}} = {s_{BB}}(t)
\end{equation}
where ${s_{RBB}}(t)$ is the demodulation signal, ${e^{i\omega t}}$ is the R-frequency signal, ${s_{CB}}(t)$ is the modulation signal, ${s_{BB}}(t)$ is baseband signal. The band move of the demodulation for the L-complex modulation signal is shown in Fig. \ref{fig9}.

The R-complex modulation signal is demodulated by the L-frequency signal as the following formula:
\begin{equation}
\label{eq16}
{s_{RBB}}(t) = {s_{CB}}(t)*{e^{-i\omega t}} = ({s_{BB}}(t)*{e^{i\omega t}})*{e^{ - i\omega t}} = {s_{BB}}(t)
\end{equation}
where ${s_{RBB}}(t)$ is the demodulation signal, ${e^{-i\omega t}}$ is the L-frequency signal, ${s_{CB}}(t)$ is the modulation signal, ${s_{BB}}(t)$ is baseband signal. The band move of the demodulation for the R-complex modulation signal is similar and omitted.

\begin{figure}
\includegraphics[width=5in]{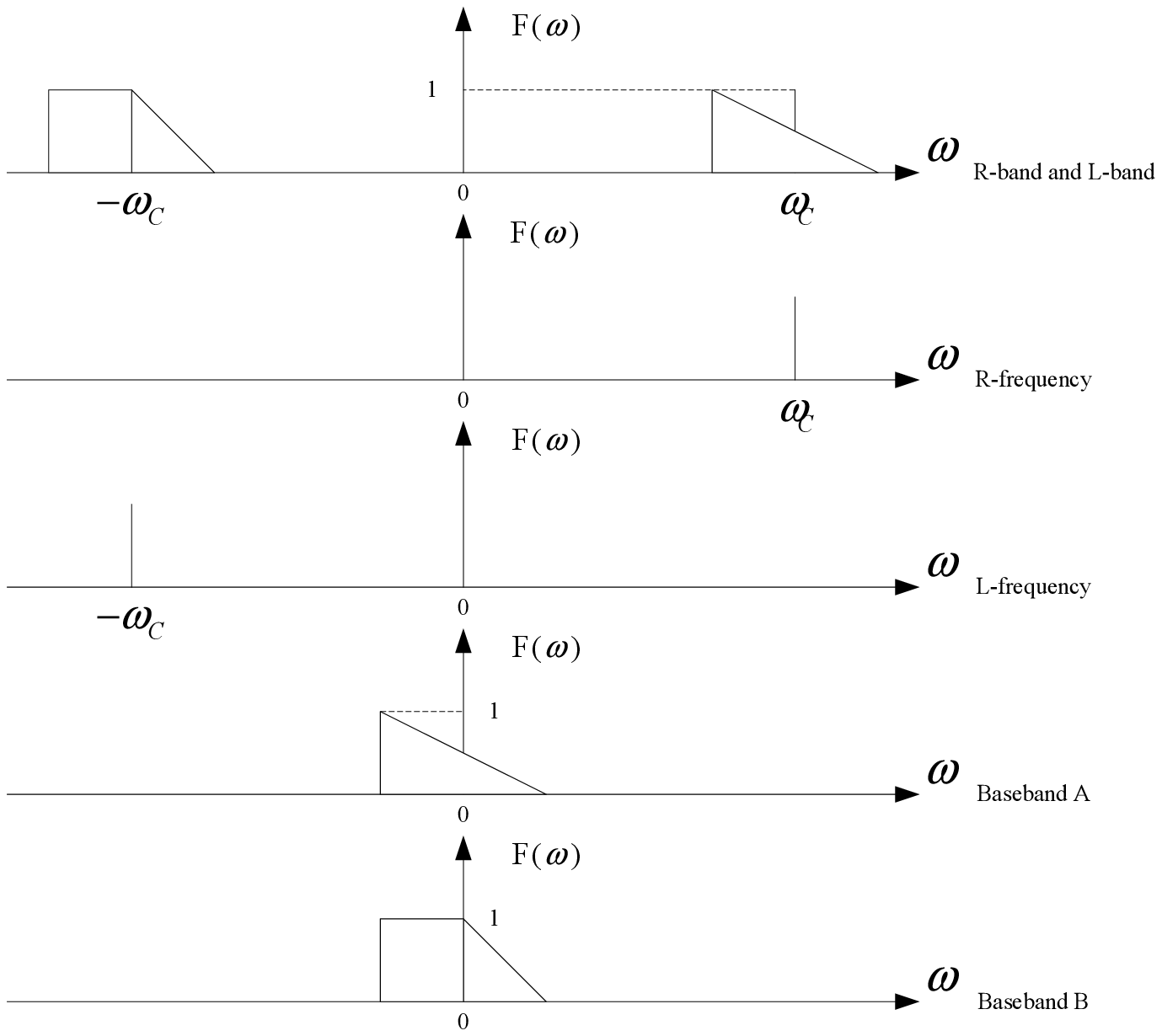}\\
\caption{Band move of demodulating two different information.}\label{fig10}
\end{figure}

The different information on the L-band and R-band can be demodulated separately, as shown in Fig. \ref{fig10}.

\subsection{Essence of complex-carrier modulation and demodulation}
As mentioned earlier, in essence, the principle of complex-carrier modulation and demodulation are the same, they are band moves. So in this paper, we use a single word ``band-move'' instead of the modulation and demodulation.

Furthermore, band-move is a transformation, it is the rotational speed transform in time domain, the move transform in frequency domain. Like the reference system transform, we can see, the frequency is a relative value, its value is related to the reference frequency, as the following formula:
\begin{equation}
\label{eq17}
{e^{i{\omega ^*}t}} = {e^{i\omega t}}*{e^{i{\omega _c}t}} = {e^{i(\omega  + {\omega _c})t}}
\end{equation}
where ${e^{i{\omega ^*}t}}$ is the frequency after transform, ${e^{i\omega t}}$ is the frequency before transform, ${e^{i{\omega _c}t}}$ is the reference frequency.

Thus, a ``negative frequency'' can become a ``positive frequency'' after the band-move transform, which also confirms, on the other hand, that the ``negative frequency'' has a physical meaning, and the ``sign'' of a frequency is related to the reference frequency.

The transformation has the following properties:

1: The additive property, one band-move is equivalent to the sum of several band-moves, as the following formula:
\begin{equation}
\label{eq18}
{s_{CB}}(t)={s_{BB}}(t)*{e^{i{\omega _1}t}}*{e^{i{\omega _2}t}}={s_{BB}}(t)*{e^{i({\omega _1} + {\omega _2})t}}
\end{equation}
where ${s_{CB}}(t)$ is the band-move signal, ${s_{BB}}(t)$ is the original signal.

2: The commutative property, the result of several band-moves has nothing to do with the orders, as the following formula:
\begin{equation}
\label{eq19}
{s_{CB}}(t) = {s_{BB}}(t)*{e^{i{\omega _1}t}}*{e^{i{\omega _2}t}} = {s_{BB}}(t)*{e^{i{\omega _2}t}}*{e^{i{\omega _1}t}}
\end{equation}
where ${s_{CB}}(t)$ is the band-move signal, ${s_{BB}}(t)$ is the original signal.

So the transformations of band-move make up a continuous Abel group.

\section{Circularly polarized electromagnetic signal}
From this section, we will see that the circularly polarized electromagnetic signals are exactly the complex-frequency signals, the right-hand and left-hand circularly polarized electromagnetic signals correspond to the R-frequency and L-frequency signals.

As we all know, light is electromagnetic wave, and the right-hand circularly polarized light wavefront is described as the following formula:
\begin{equation}
\label{eq20}
\overrightarrow {\rm{E}} = A*{e^{i(\omega t - 2\pi *x/\lambda  + \phi )}}
\end{equation}
where $\overrightarrow {\rm{E}}$ is electric vector, $A$ is the electric intensity, $\omega$ is the angular frequency, $x$ is the propagate distance, $\lambda$ is the wavelength, $\phi$ is the initial phase.

Assume we observed this light at a certain place, that means $x$ is a constant, so $- 2\pi *x/\lambda  + \phi$ is a constant, it can be seen as a initial phase $\phi^*$, so the (\ref{eq20}) can be change to the following formula:
\begin{equation}
\label{eq21}
\overrightarrow {\rm{E}}=A*{e^{i(\omega t-2\pi*x/\lambda+\phi)}}=A*{e^{i\omega t+\phi^*}}
=A*cos(\omega t+\phi^*)+i*A*sin(\omega t+\phi^*)
=A*{e^{i\phi^*}}*{e^{i\omega t}}
\end{equation}
It is a R-frequency $e^{\i\omega t}$ which has a initial phase $\phi^*$ and a non-normalized amplitude $A$. Also we can see it contains a real part $A*cos(\omega t+\phi^*)$ and a imaginary part $A*sin(\omega t+\phi^*)$, so the right-hand circularly polarized light is a complex-frequency signal, and the real part and imaginary part are orthogonal in space, that means we can send a complex-frequency signal by a circularly polarized signal.

\begin{figure}
\includegraphics[width=5in]{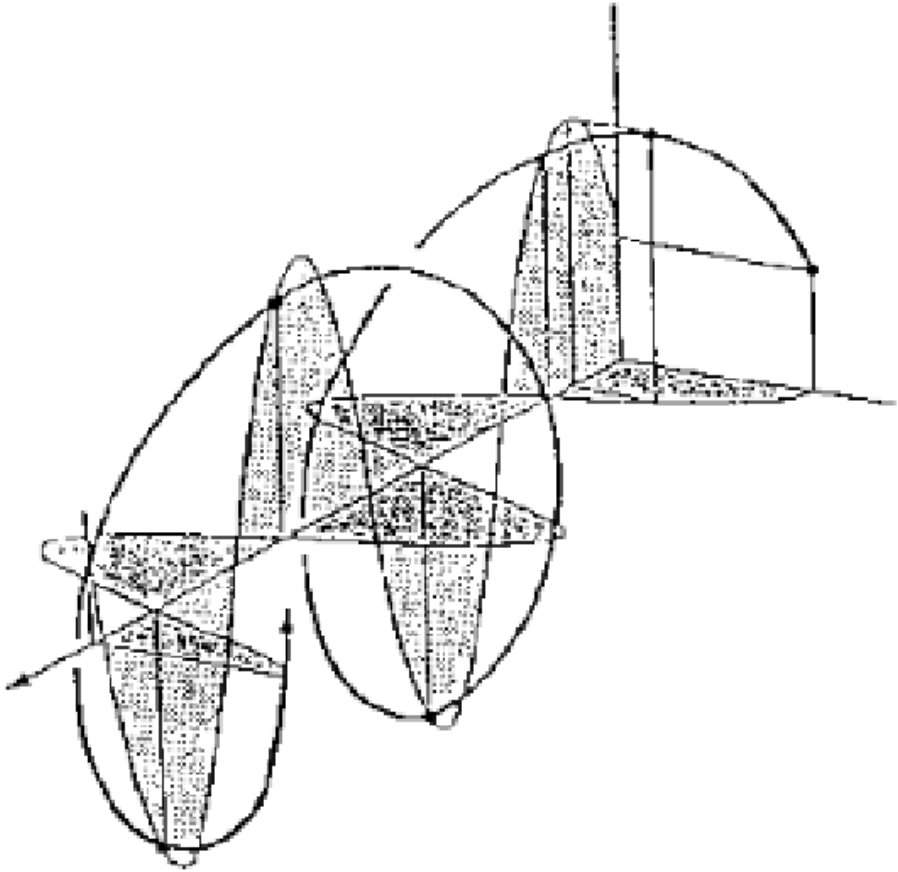}\\
\caption{Circularly polarized electromagnetic signal.}\label{fig11}
\end{figure}

The right-hand circularly polarized light is shown in Fig. \ref{fig11}, it is the same as the Fig. \ref{fig3}.

Similarly, we can see the left-hand circularly polarized light is the R-frequency signal. Furthermore, the linearly polarized electromagnetic signal is the real-frequency signal.

So we can send out the R-band and L-band signals by the right-hand and left-hand circularly polarized electromagnetic signals.

By the way, formula (\ref{eq20}) is a solution of Schrodinger equation, as shown in Fig. \ref{fig11}, we can see, here ``i'' has a clearly physical meaning, it means orthogonal in space, not imaginary.

\section{Conclusions}
In summary, we draw the following conclusions:

1: Negative spectrum exists, but is wasted.

2: A complete description of a frequency signal is a rotating complex-frequency signal, in a complete description, R-frequency signal and L-frequency signal are two distinguishable and independent frequency signals.

3: Current real-carrier modulation don't distinguish between R-frequency signal and L-frequency signal, so R-band and L-band are both occupied.

4: Current real-carrier demodulation only receive one band, so half of the energy will be lost.

5: Image frequency is caused by the real-carrier modulation and demodulation.

6: Complex-carrier modulation use the complex-frequency signal as a carrier signal, so it can carry different information on L-band and R-band, the spectrum resources are fully used.

7: Complex-carrier demodulation can receive the whole signal energy, so the channel capacity is higher.

8: Frequency is a relative concept, it is relative to the selection of reference frequency.

9: Complex-carrier modulation and demodulation is essentially a transformation, make up a continuous Abel group.

10: Circularly polarized electromagnetic signal is the complex signal, while linearly polarized signal is the real signal.

In fact, the conception of complex-carrier is just like the subcarrier in OFDM, but expanded to the RF range.

Due to the complex-carrier modulation and demodulation technology can make full use of the spectrum, and do not discard the signal energy, therefore, facing with the scarce of the spectrum resources, we sure that the complex-carrier modulation and demodulation technology will become the mainstream of the next-generation communication technology. For example, in the current LTE technology, if the two code-words are modulated by the L-band and R-band separately, then clearly the channel capacity will be greatly improved. For another example, if up-link and down-link are modulated by the L-band and R-band separately, then up-link and down-link can communicate using the 'same' frequency at the same time.

\end{document}